\documentclass{amsart}
\usepackage{amsmath}
\usepackage{setspace}

\newtheorem{theorem}{Theorem}[section]

\newtheorem{lemma}[theorem]{Lemma}

\theoremstyle{definition}

\newcommand{\ep}{\varepsilon}

\renewcommand{\l}{\lambda}
\newcommand{\ra}{\rightarrow}
\renewcommand{\d}{\delta}
\newcommand{\g}{\gamma}

\title[Nonlocal Gordon]{Exact solitons in the nonlocal Gordon equation}

\author{Adam Chmaj \& Leszek Zabielski}
\address{Faculty of Mathematics and Information Science \\
Warsaw University of Technology\\ Pl. Politechniki 1 \\
00-661 Warsaw \\ Poland} \email{A.Chmaj@mini.pw.edu.pl \\
lz@mini.pw.edu.pl}

\begin{document}

\begin{abstract}
We find exact monotonic solitons in the nonlocal Gordon equation
$u_{tt}=J*u-u-f(u)$, in the case $J(x)=\frac{1}{2}e^{-|x|}$. To this
end we come up with an inverse method, which gives a representation
of the set of nonlinearities admitting such solutions. We also study
$u^{(iv)}+\l u'' -\sin u=0$, which arises from the above when we
write it in traveling wave coordinates and pass to a certain limit.
For this equation we find an exact $4\pi$-kink and show the
non-existence of $2\pi$-kinks, using the analytic continuation
method of Amick and McLeod.
\end{abstract}

\maketitle

\section{Introduction}
\setcounter{equation}{0}

The sine-Gordon equation
\begin{equation}
\label{lsg} u_{tt} = u_{xx}-\sin u
\end{equation}
is one of the most popular PDEs. It was studied in connection with
pseudospherical surfaces by B\"acklund in the 19th century
\cite{tu}, it is completely integrable and appears in many physical
models. For example, the solitons solutions $u(x,t)=U(x-c t)$,
$c^2<1$, $U(\mp \infty)=0$, $U(\pm \infty)=2\pi$, which have the
exact form
\begin{equation}
u(x,t)=  4 \arctan \exp \left( \pm \frac{x-ct}{\sqrt{1-c^2}}\right),
\label{es}
\end{equation}
obey relativistic dynamics and may be treated as particles in a
field theory \cite{ps}.

The nonlocal sine-Gordon equation
\begin{equation}
u_{tt}=J*u-u-\sin u, \label{nsg}
\end{equation}
where $J*u(x)\equiv \int_R J(x-y)u(y)dy$, $\int_R J(x)dx=1$ and
$J(-x)=J(x)$, arises e.g., in Josephson tunnel junctions made of
high temperature superconductors \cite{as,aekm}. Its discrete
counterpart
\begin{equation}
\label{dsg}
\ddot{u}_n = u_{n+1}-2 u_n +u_{n-1} -\sin u_n .
\end{equation}
has been used to model dislocations in crystals, as it represents
the motion of a chain of gravity physical pendulums coupled by
linear springs.

The study of solitons of (\ref{nsg}) and (\ref{dsg}) is of immediate
interest. So far, only exact and numerical solutions have been
found, of different type than (\ref{es}) \cite{pk}. In \cite{as}, it
was discovered that $u(x-ct)=4\pi \arctan(x-ct)$, $c^2 =1$, is a
solution of the sine-Hilbert equation
\begin{equation}
u_{tt}=\frac{1}{\pi} p.v. \int_R \frac{u_x (y)}{y-x}dx-\sin u,
\label{sH}
\end{equation}
which arises as a limit of (\ref{nsg}) with an appropriate $J$. This
$4\pi$-kink carries two magnetic flux quanta \cite{as}. In
\cite{aekm} the authors studied the case with
$J(x)=\frac{1}{2}e^{-|x|}$ and stated a Hypothesis 1
\cite[p.406]{aekm}, that there are no $2\pi$-kinks for $0<c^2 <1$.
Instead, they numerically determined that solitons develop periodic
oscillations around $0$ and $2\pi$ at $\mp \infty$, which has been
interpreted as a Cherenkov radiation phenomenon. We mention that
homoclinic asymptotically periodic waves were first formally
\cite{prg} and then rigorously \cite{at} constructed for a fourth
order KdV equation. Also in \cite{aekm}, the authors found numerical
evidence for $4\pi$- and $6\pi$-kinks for some values of $c^2$.

In this note, we approach the problem of existence or nonexistence
of monotonic solitons of (\ref{nsg}) from a different point of view.
We replace $\sin u$ with a general $f(u)$, as we think that this
problem is not structurally stable with respect to the nonlinearity.
Namely, we study traveling wave solutions \mbox{$u(x-c t)$} to
\begin{equation}
\label{snsg} u_{tt}=  \frac{1}{\ep^2}(J_\ep * u - u) - f(u),
\end{equation}
where $J_\ep (x)\equiv \frac{1}{\ep} J(\frac{x}{\ep})$. It is
natural to consider this scaling, as formally with $\ep\to 0$ such a
solution approaches a traveling wave of (\ref{lsg}). We take
$J(x)=\frac{1}{2}e^{-|x|}$ as in \cite{aekm}. $u(y)=u(x-ct)$ then
satisfies
\begin{equation}
\ep^2 c^2 u^{(iv)} +(1-c^2 )u'' -f(u) +\ep^2 f(u)''=0. \label{4}
\end{equation}
Let $u'>0$, $W(s)=\int_{-a}^s f$ and
$z(s)=\frac{1}{2}[u'(u^{-1}(s))]^2 +\frac{W(s)}{c^2}$. We show below
that (\ref{4}) is then equivalent to
\begin{equation}
W=c^2 z -\frac{z+\frac{1}{2} \ep^2 c^2 z'^2}{(\frac{1}{c^2}+\ep^2
z'')^2}. \label{r}
\end{equation}
Thus $f$ admits a monotonic soliton only if $W$ is in the image of
the mapping (\ref{r}). Moreover, from (\ref{r}) we see that for
every $f$ there exists a close $f_\ep$ which admits a monotonic
soliton (Theorem \ref{theorem1} below). Likely a similar result
holds for (\ref{nsg}) with a general $J\geq 0$ and for (\ref{dsg}).
In \cite[p. 251]{bz}, the authors speculated that (\ref{snsg})
admits a soliton connecting consecutive stable roots of any balanced
bistable $f$, at least for small $\ep >0$. We think that one cannot
expect much more beyond Theorem \ref{theorem1}, and a proof of the
aforementioned Hypothesis 1 in \cite{aekm} would disprove this
conjecture. Some methods were introduced for showing the
nonexistence of monotonic heteroclinic solutions of a third order
equation bearing some similarity to (\ref{4}) \cite{t,hm,am,in},
however, they do not seem to be easily applicable here.

In \cite{s} the author discovered that $u(y)=\tanh (k y)$ is a
solution of
\[ c^2 u''(y)=d(u(y+1)-2u(y)+u(y-1))-f(u(y)) , \]
where
\begin{equation}
W(u)=(c^2 k^2 -d)u^2 -\frac{c^2 k^2}{2} u^4 -\frac{d}{\sinh^2 k} \ln
(\cosh^2 k -u^2 \sinh^2 k ). \label{Wd}
\end{equation}

In Section 2, we derive (\ref{r}) and discuss its consequences,
e.g., Theorem \ref{theorem2}, in which we construct a $W$ with any
number of arbitrarily spaced wells, admitting solitons connecting
the outermost ones. This generalizes (\ref{Wd}) in a nontrivial way,
since $W$ in (\ref{Wd}) has only three equal depth wells for an
appropriate choice of parameters. In Section 3, we consider
(\ref{4}) without the term $\ep^2 (f(u))''$, but with $f(u)=\sin u$.
We find an exact $4\pi$-kink (Theorem \ref{theorem3}) and use the
analytic continuation method from \cite{am} to show that such a
simpler equation admits no $2\pi$-kinks (Theorem \ref{theorem4}).

\section{The  main result}
\setcounter{equation}{0}

Let $f$ be a balanced bistable nonlinearity. To be more precise, we
assume that
\begin{equation}
f\in C^1 ,~~ f(\pm a)=f(0)=0,~~ f|_{(-a,0)}>0, ~~f|_{(0,a)}<0,
~~f'(\pm a)>0,~~ \int_{-a}^a f= 0. \label{as}
\end{equation}
We reach (\ref{r}) as follows. Let $\l_c = \frac{1-c^2}{c^2}$ and
$f_c =\frac{1}{c^2}f$. Then ({\ref{4}) becomes
\begin{equation}
\label{odenlsg} \ep^2 u^{(iv)} +\lambda_c u'' -f_c(u) +\ep^2
f_c(u)''=0
\end{equation}
The reduction to a second order equation that follows is similar in
spirit to that in \cite{pt} applied to the equation $-\gamma
u^{(iv)} +u''-f(u)=0$, though the calculation here is trickier and
the end result is different than the one in \cite{pt}. First we find
the first integral of (\ref{odenlsg}). Multipling (\ref{odenlsg}) by
$u'$ and integrating from $-\infty$ to $y$ leads to
\begin{equation}
\ep^2 u''' u' - \frac{1}{2} \ep^2 u''^2 +\frac{1}{2} \l_c u'^2
-W_c(u)+\ep^2 \int_{-\infty}^y (f_c(u))''u' =0 ,\label{fi1}
\end{equation}
where $W_c=\frac{1}{c^2}W$. The integral term in (\ref{fi1}) is
handled in the following way. First we integrate twice by parts:
\begin{eqnarray}
\nonumber
 \ep^2 \int_{-\infty}^y (f_c(u))''u' &= & \ep^2 f_c'(u)u'^2 - \ep^2
 \int_{-\infty}^y
 (f_c(u))'u'' \\ \label{fi2}
& = &  \ep^2 f_c'(u)u'^2 -\ep^2 f_c(u)u'' +\ep^2 \int_{-\infty}^y
f_c(u)u''' .
\end{eqnarray}
Then we integrate (\ref{odenlsg}) and substitute
\[ \ep^2 u'''=\int_{-\infty}^y f_c(u) -\l_c u' - \ep^2 (f_c(u))' \]
into the integral term in (\ref{fi2}):
\begin{equation}
\ep^2 \int_{-\infty}^y f_c(u)u'''=\frac{1}{2} [\ep^2 u'''+\l_c u'+\ep^2
(f_c(u))']^2 -\l W_c(u)-\frac{1}{2} \ep^2 f_c(u)^2 .\label{fi3}
\end{equation}
Substituting (\ref{fi3}) into (\ref{fi2}), and (\ref{fi2}) into
(\ref{fi1}), we obtain the first integral of (\ref{odenlsg}):
\[ \ep^2 u'''u' -\frac{1}{2} \ep^2 u''^2 + \frac{1}{2}\l_c u'^2
-W_c(u)+\ep^2 f_c'(u)u'^2 -\ep^2 f_c(u)u'' \]
\begin{equation}
+ \frac{1}{2} [\ep^2 u''' +\l_c u' +\ep^2 (f_c(u))']^2 -\l_c
W_c(u)-\frac{1}{2} \ep^2 f_c(u)^2 =0. \label{fi4}
\end{equation}
We consider only solutions $u$ such that $u'>0$. Let $s= u(y)$ and
$y(s)= u^{-1}(s)$. We reduce the order in
(\ref{fi4}) with the substitution $v(s)=[u'(y(s))]^2$. Note that
$v'(s)= 2 u''(y(s))$ and $v''(s)= \frac{2 u'''}{u'}$. We get:
\[ \frac{1}{2} \ep^2 v''v-\frac{1}{8}\ep^2 v'^2 +\frac{1}{2}\l_c v
+\ep^2 f_c'(s)v -\frac{1}{2}\ep^2 f_c(s)v' \]
\[ +\frac{1}{2} [\frac{1}{2} \ep^2 v'' \sqrt{v}+\l_c \sqrt{v}
+\ep^2 f_c'(s)\sqrt{v}]^2 -\frac{1}{2} \ep^2 f_c(s)^2 -(1+\l_c
)W_c(s)=0 .\]
It is probably remarkable that this equation can be
simplified with the substitution $z=\frac{1}{2}v+W_c(s)$, to give
\[
-\frac{1}{2}\ep^2 z'^2 +(z-W_c(s))(1+\l_c +\ep^2 z'')^2 -(1+\l_c
)z=0, \] or,
\begin{equation}
-\frac{1}{2}\ep^2 c^2 z'^2 +(c^2 z-W(s))(c^{-2} +\ep^2 z'')^2 -z=0,
\label{z}
\end{equation}
which decouples the nonlinearity from the solution and yields
(\ref{r}).

A positive solution of (\ref{z}) with the boundary conditions $z(\pm
a)=z'(\pm a)=0$ yields a solution of (\ref{4}) with $u(\pm
\infty)=\pm a$, if in addition
\begin{equation} y(u)=\int_0^u y'(s)ds
= \int_0^u \frac{ds}{\sqrt{v(s)}} \ra \pm \infty ~~ {\rm as} ~~ u
\ra \pm a. \label{back}
\end{equation}
Let us assume that $\lim_{s\ra a}z''(s)$ exists. Denote it by $L$.
As in \cite{pt}, after dividing both sides of (\ref{z}) by $z$ and
passing to the limit, we get
\[ \Bigl( c^2 -\lim_{s\ra a}\frac{W(s)}{z(s)}\Bigr) (c^{-2}+\ep^2 L)^2
= 1+ \frac{1}{2}\ep^2 c^2 \lim_{s \ra a} \frac{z'^2 (s)}{z(s)}. \]
After using l'H\^{o}pital's rule we get
\begin{equation}
\Bigl( 1-\frac{f'(a)}{Lc^2} \Bigr) (c^{-2} +\ep^2 L)=1 , \label{f'}
\end{equation}
or
\[ \ep^2 c^2 L^2 +(1-c^2 -\ep^2 f'(a))L -c^{-2}f'(a)=0 ,\]
which has the solutions
\[ L_{\pm} = \frac{1}{2\ep^2 c^2}\Bigl[ -(1-c^2 -\ep^2 f'(a)) \pm
\sqrt{(1-c^2 -\ep^2 f'(a))^2 +4 \ep^2 f'(a)} \Bigr] .\] After
squaring, $\frac{1}{2}v''(a) =L_+ -\frac{f'(a)}{c^2}>0$ is
equivalent to $1>1-c^2$, hence for some $K>0$, $v(s)\sim K(a-s)^2$
as $s\ra a$. As a similar argument applies at $s\ra -a$, we see that
(\ref{back}) is satisfied.

As an application of (\ref{r}), we get the following results.

\begin{theorem}
Let $f_0$ satisfy (\ref{as}). For any $0<c^2<1$, there is an $\ep(c)
>0$, such that for $0<\ep < \ep(c)$ there are solution pairs $(u_\ep
,f_\ep )$ of (\ref{4}), in the sense that $u_\ep$, such that $u_\ep
'>0$ and $u_\ep (\pm \infty)=\pm a$, satisfies (\ref{4}) with
$f_\ep$. Moreover, $(u_\ep ,f_\ep ) \ra (u_0 ,f_0 )$ as $\ep \ra 0$,
where $u_0$ satisfies $(1-c^2) u_0'' -f_0 (u_0)=0$. \label{theorem1}
\end{theorem}

{\it Proof.} Let $z_0 (s)=\frac{1}{2}[u_0 '(u_0^{-1} (s))]^2
+\frac{W_0 (s)}{c^2}$. For $0<\ep <\ep(c)=\sqrt{ -\frac{1}{c^2 \min
z_0 ''(t)}}$, we can define $W_\ep =c^2 z_0 -\frac{z_0 +\frac{1}{2}
\ep^2 c^2 {z_0 '}^2 }{(\frac{1}{c^2} +\ep^2 z_0 '')^2}$ and
$\frac{1}{2}v_\ep =z_0 -\frac{W_\ep}{c^2}$. Let $L_0 =\lim_{s \ra a}
z_0 '' (s)$. From (\ref{f'}), $f_\ep '(a)=L_0 \frac{1-c^2 +\ep^2
L_0}{c^{-2}+\ep^2 L_0}>0$, thus $v_\ep ''(a)>0$ and $u_\ep$ defined
by $y(u_\ep )=\int_0^{u_\ep} \frac{ds}{\sqrt{v_\ep (s)}}$ is a
solution corresponding to $f_\ep =W_\ep '$. $\Box$

\vspace{5mm}

{\it Remark.} Recall that if $f_0$ is multistable, e.g., $f_0
(u)=\sin u$, then $(1-c^2) u_0'' -f_0 (u_0)=0$ has only solitons
connecting nearest stable zeroes of $f_0$. However, this is not the
case for (\ref{4}), as was e.g., determined numerically in
\cite{aekm}. Indeed, let $z_0$ correspond to $f_0 (u)=u^3 -u$, i.e.,
$z_0 (t) =\frac{(t^2-1)^2}{4c^2 (1-c^2)}$. It can be verified that
for $\ep_t (c)=\sqrt{(1-c^2)(1-|c|)}$, $W_{\ep_t (c)}$ is a
triple-well function with wells of equal depth, i.e., $W_{\ep_t
(c)}(\pm 1)=W_{\ep_t (c)}(0)=0$. For $\ep_t (c)<\ep <\ep (c)$,
$W_\ep (0)<0$. This example is similar to (\ref{Wd}) \cite{s}.
However, (\ref{r}) enables us to go further.

\begin{theorem}
Let $a_1 , \ldots, a_n$ be an increasing sequence. For any
$0<c^2<1$, there exists a multi-well potential ${\tilde W}$, such
that ${\tilde W} (a_k )=0$, $k=1,\ldots,n$, ${\tilde W}>0$
elsewhere, and a soliton solution ${\tilde u}'>0$ of (\ref{4}) with
${\tilde f}={\tilde W}'$, such that ${\tilde u}(-\infty)=a_1$,
${\tilde u}(+\infty)=a_n$. \label{theorem2}
\end{theorem}

{\it Proof.} Let $\bar z$ have zeroes at $\pm m=\min \{ a_2 -a_1 ,
a_n -a_{n-1} \}$ and ${\bar \ep}_t$ be the value corresponding to
$W_{{\bar \ep}_t}$ being a triple well function, as in the above
Remark.

We paste and glue. Cover $a_2 ,\ldots,a_{n-1}$ with disjoint closed
intervals $I_2 ,\ldots,I_{n-1}$, each of the same length less than
$2m$. On each $I_k =[i_k ,j_k]$, $k=2,\ldots,n-1$, let ${\tilde z}
(t)= {\bar z} (t-i_k - \frac{|I_k|}{2})$. Between those intervals
extend ${\tilde z}$ smoothly and sufficiently close to a constant.
Without loss of generality, let $[a_1 ,a_2 ]$ be shorter than
$[a_{n-1},a_n ]$. On $[a_1 ,i_2]$ let ${\tilde z} (t)= {\bar z} (t
-a_1 -m )$, on $[a_n -(i_2 -a_1 ),a_n]$ let ${\tilde z} (t)= {\bar
z} (t -a_n +m)$. On $[j_{n-1},a_n -(i_2 -a )]$ extend ${\tilde z}$
smoothly and sufficiently close to a constant.

Then ${\tilde W}$ defined by (\ref{r}) with $z={\tilde z}$ and
$\ep={\bar \ep}_t$ has the required properties. The soliton $\tilde
u$ is obtained as in the Proof of Theorem 2.1. $\Box$

\section{An asymptotic limit equation with sine}
\setcounter{equation}{0}

In \cite{aekm} the authors suggested considering solutions of
(\ref{4}) of the form $U(y)=u(\sqrt{\ep} y)$. $U$ is then a solution
of
\begin{equation} c^2 U^{(iv)}+\frac{1-c^2}{\ep} U''-f(U) +\ep
f(U)''=0 .\label{4s}
\end{equation}
Assuming $c^2 \ra 1$, $\ep \ra 0$ and $\frac{1-c^2}{\ep}\ra \l
>0$, we obtain the simpler
\begin{equation} U^{(iv)}+\l U'' -f(U)=0 .\label{4a}
\end{equation}
First we present a similar result as the one for (\ref{sH}) in
\cite{as}.

\begin{theorem}
For $\l =\frac{\sqrt{3}}{32}$
\begin{equation}
U(y)=8\arctan \bigl( \exp \bigl( \frac{\sqrt{2}}{\sqrt[4]{3}}
y\bigr) \bigr) \label{ex}
\end{equation}
is an exact $4\pi$-kink solution of (\ref{4a}) with $f(U)=\sin U$.
\label{theorem3}
\end{theorem}

{\it Remark.} In \cite[p.409]{aekm}, the authors' numerical results
give the first such $4\pi$-kink at $\l \approx 1.155$.

{\it Proof of Theorem \ref{theorem3}.} We show how we found
(\ref{ex}). Consider only solutions such that $U'>0$ and let
$v(s)=[U'(U^{-1}(s))]^2$, as in Section 2. Using the first integral
of (\ref{4a}), we get
\begin{equation}
vv''-\frac{v'^2}{4} +\l v -2W(t)=0. \label{za}
\end{equation}
We try if $v(s)=1-\cos \frac{s}{2}$ is an exact solution. For $\l
=\frac{1}{4}$ it indeed satisfies (\ref{za}) with $W(s)=\frac{3}{64}
(1-\cos s )$, thus ${\bar U}(y)=8\arctan
(\exp(\frac{\sqrt{2}}{4}y))$ determined from
\[ y({\bar U})=\int_{2\pi}^{\bar U} \frac{dt}{\sqrt{v(s)}}=
\int_{2\pi}^{\bar U} \frac{ds}{\sqrt{2} \sin \frac{s}{4}} =
\frac{4}{\sqrt{2}} \ln \tan \bigl( \frac{{\bar U}}{8} \bigr) \]
satisfies
\[ {\bar U}^{(iv)}+\frac{1}{4} {\bar U}'' -\frac{3}{64} \sin {\bar U} =0 .\]
Therefore $U(y)={\bar U}(\frac{y}{\sqrt[4]{\frac{3}{64}}})$
satisfies (\ref{4a}) for $\l =\frac{\sqrt{3}}{32}$ and is given by
(\ref{ex}). $\Box$

However, there are no $2\pi$-kinks for (\ref{4a}) with $f(U)=\sin
U$. Since we will be working in the complex plane, it is convenient
to switch the notation, so that $U\equiv U(x)$.

\begin{theorem}
Let $\l >0$ and $f(U)=\sin U$. There exist no solutions $U$ of (\ref{4a}),
such that $U(-\infty)=0$, $U(+\infty)=2\pi$. \label{theorem4}
\end{theorem}

{\it Proof.} To simplify the notation, let $w(x)=-\pi+U(x)$. Then
$w$ satisfies
\begin{equation}
w^{(iv)}+\l w'' +\sin w=0 .\label{w}
\end{equation}
In Lemmas \ref{mon} and \ref{asym} below we show that $w'>0$ and $w$
is odd. With this at hand, we can adapt the analytic continuation
approach method in \cite{am}, where the authors established a
similar result for the equation $\ep w'''+w'-\cos w=0$, see also
\cite{am4,jtm} for extensions to some equations with polynomial
nonlinearities. It is reminiscent of Painlev\'e transcendents and
some properties of the inverse scattering theory. In this context,
the inspiration might had also been drawn from some beyond all
orders asymptotics results, e.g., \cite{sk,prg}.

We argue by contradiction. First the solution $w$ is analytically
continued to $w(z)$, $z\in C$, in such a way that it retains some
properties of $w_0 (z)=-\pi +4\arctan \exp \sqrt{\l} z$, which
solves $\l w_{xx}+\sin w=0$. Then, what we think is the main idea of
the method, is that these properties are incompatible with the
structure of (\ref{w}) near the singularity of $w$. We will also see
the limitation of this approach, namely, that at this stage there is
compatibility with the structure of the complete nonlocal equation
(\ref{4}) with $f(u)=-\sin u$.

Recall that $\arctan z=\frac{1}{2i}\log \frac{1+iz}{1-iz}$. Using
this representation, if we define $\log$ to have a branch cut along
the positive real axis, then $w_0$ has branch cuts along the lines
$z=x+i \sqrt{\l} (\frac{\pi}{2}+k\pi )$, $x<0$, $k\in Z$. Moreover,
if $w_0 (z) =p_0 (x,y)+iq_0 (x,y)$, then $p_0 = \pi$ for $z
=x+i\sqrt{\l} \frac{\pi}{2}$, $x>0$, $p_0 =0$ for $z=iy$, $0\leq y
<\sqrt{\l}\frac{\pi}{2}$, $p_0 =2\pi$ for $z=iy$,
$\sqrt{\l}\frac{\pi}{2}< y \leq \sqrt{\l}\pi$ and $p_0 (x,y)=2\pi
-p_0 (x,\sqrt{\l}\pi -y)$, $q_0 (x,y)=q_0 (x,\sqrt{\l}\pi -y)$ for
$x>0$, $0<y\leq \sqrt{\l}\frac{\pi}{2}$.

Let $m_1$ denote the positive root of $m^4 +\l m^2 -1=0$. Below we
show that $w$ can be extended to $w(z)$, where $z=x+iy$, $x\geq 0$,
$0\leq y\leq \frac{\pi}{m_1}$, $y\neq  \frac{1}{2} \frac{\pi}{m_1}$,
which is a solution of $w_{xxxx} +\l w_{xx}+\sin w=0$, in such a way
that if $w(z)=p(x,y)+iq(x,y)$, then
\begin{equation}
\begin{array}{l}
p (x,\frac{1}{2} \frac{\pi}{m_1})= \pi ,~~x>0, \\
q(x,0) =0,~~x>0,\\
p (0,y)=0,~~ 0\leq y <\frac{1}{2}\frac{\pi}{m_1},\\
p (0,y)=2\pi ,~~  \frac{1}{2} \frac{\pi}{m_1}< y \leq
\frac{\pi}{m_1},\\
p (x,y)=2\pi -p (x,\frac{\pi}{m_1} -y)~~{\rm and} \\q (x,y)=q
(x,\frac{\pi}{m_1} -y)~~{\rm for}~~ x>0, ~~0<y\leq
\frac{1}{2}\frac{\pi}{m_1}.
\end{array} \label{pq}
\end{equation}
At $z=i\frac{1}{2}\frac{\pi}{m_1}$, $w$ has a singularity. We extend
$w$ as a solution of the equation to the left hand strip:
\[ \begin{array}{l}
p(-x,y)=-p(x,y),~~0\leq y <\frac{1}{2}\frac{\pi}{m_1},\\
p(-x,y)=4\pi - p(x,y),~~  \frac{1}{2} \frac{\pi}{m_1}< y \leq
\frac{\pi}{m_1},\\
q(-x,y)=q(x,y),~~ 0\leq y \leq \frac{\pi}{m_1} .
\end{array}\]
Thus $w$ is analytic in the whole strip $S=\{ (x,y):0\leq y \leq
\frac{\pi}{m_1}\}$, with the exception of the half line
$z=x+i\frac{1}{2} \frac{\pi}{m_1}$, $x\leq 0$, across which $p$ is
discontinuous:
\[ \lim_{y\ra \frac{1}{2} \frac{\pi}{m_1}+} p(x,y)-\lim_{y\ra
\frac{1}{2} \frac{\pi}{m_1}-} p(x,y)= 4\pi ,~~x<0 .\] Let
\begin{equation}
\begin{array} {l}
h(z)=w(z)+2i\log (z-i\frac{1}{2} \frac{\pi}{m_1}),
\end{array}\label{h}
\end{equation}
where this $\log$ has a branch cut across the negative real axis.
Since $h$ is continuous across the half line $z=x+i\frac{1}{2}
\frac{\pi}{m_1}$, $x\leq 0$, it is also analytic in $S$, with the
exception of the point $(0,\frac{1}{2} \frac{\pi}{m_1})$. Let $h=p_1
+iq_1$. Since $p_1 =p-2\arg (z-i\frac{1}{2} \frac{\pi}{m_1})$ and
$p$ is bounded, $p_1$ is also bounded. From Big Picard Theorem, $h$
cannot have an essential singularity at $(0,\frac{1}{2}
\frac{\pi}{m_1})$. An elementary calculation also shows that if
$p_1$ is bounded, the singularity of $h$ cannot be a pole. Thus $h$
is analytic. Substituting (\ref{h}) into (\ref{w}), we obtain a
contradiction, since $w_{xx}$ and $\sin w$ are both of order
$O((z-i\frac{1}{2} \frac{\pi}{m_1})^{-2})$, while $w_{xxxx}$ is of
order $O((z-i\frac{1}{2} \frac{\pi}{m_1})^{-4})$.

{\it Remark.} There is no such contradiction if we consider
(\ref{4}) in place of (\ref{w}), as $(\sin w)_{xx}$ is also of order
$O((z-i\frac{1}{2} \frac{\pi}{m_1})^{-4})$. $\Box$

To complete the proof, we prove the aforementioned lemmas and
construct the analytic continuation of $w(x)$ with the needed
properties.
\begin{lemma}
$w'>0$. \label{mon}
\end{lemma}
{\it Proof.} First we need to show that $w',w'',w'''\ra 0$ as $x\ra
\pm \infty$. (\ref{w}) can be written as
\begin{equation} w''=(\l
+1)J*w-(\l+1)w +J*\sin w ,\label{conv1}
\end{equation}
where $J(x)=\frac{1}{2}e^{-|x|}$. We used $\lim_{x\ra \pm\infty}
e^{-|x|}w'''(x)=0$, which is not hard to get from (\ref{w}). From
(\ref{conv1}) and Lebesgue Dominated Convergence Theorem, $w''\ra 0$
as $x\ra \pm \infty$. Differentiating (\ref{conv1}) we get
\begin{equation} w'''=(\l
+1)J'*w-(\l+1)w' +J'*\sin w,\label{conv2}
\end{equation}
which can be written as
\[ w'=-(\l +2)J'*w -(\l +1) J*J'*w -J*J'*\sin w ,\]
thus $w' \ra 0$ as $x\ra \pm \infty$ and $w''' \ra 0$ as $x\ra \pm
\infty$ from (\ref{conv2}).

Now we can use the first integral of (\ref{w})
\[ w'''w'-\frac{1}{2} w''^2 +\frac{\l}{2} w'^2 =1+\cos w ,\]
from which we see that we can have $w'=0$ only at points at which
$w''=0$ and $1+\cos w=0$. If $w'=0$ at such a point, then to satisfy
$w(\pm \infty)=\pm \pi$, $w$ must be at a local maximum or minimum
at this or another such a point. From Taylor's expansion to fourth
order, also $w'''=0$ at that point, but then from uniqueness of
solutions of the initial value problem, $w$ is a constant solution,
thus reaching a contradiction. $\Box$

If we write (\ref{w}) as a system of first order equations, we see
that the critical point $(\pi,0,0,0)$ is not hyperbolic. This is in
contrast to the critical point $(\frac{\pi}{2},0,0)$ of the third
order equation in \cite{am}, which is hyperbolic, therefore some
arguments taken from \cite{am} need a bit of care here. Let $m_2$
denote the negative root of $m^4 +\l m^2-1=0$ and $m_3$, $m_4$ its
purely complex ones. Note that $m_1 +m_2 +m_3 +m_4 =0$.

\begin{lemma} $\pi-w(x)=O(e^{-m_1 x})$ as $x\ra \infty$ and
$w$ is odd. \label{asym}
\end{lemma}

{\it Proof.} If we could show that $\lim_{s\ra \pi}v''(s)$ and
$\lim_{s\ra \pi}v'''(s)$ exist in the representation (\ref{za}),
then we would calculate these limits as in (\ref{f'}), in particular
getting $\lim_{s\ra \pi}v''(s)=2m_1^2$, and then get the asymptotics
from $ x(w)-{\bar x}=\int_{w({\bar x})}^w \frac{ds}{\sqrt{v(s)}}$,
where ${\bar x}$ is sufficiently large. Such an argument was used in
\cite{pt}. However, since it is not easy to show that the limits
exist, we use the more robust method from \cite{cc}.

Let ${\bar w}=\pi -w$, $w_1=\int_x^\infty {\bar w}$,
$w_2=\int_x^\infty w_1$. Note that ${\bar w}$ and $w_1$ are
integrable from (\ref{w}) and $w',w'',w'''\ra 0$ as $x\ra  \infty$.
For small $\delta>0$ and large $x$ for which ${\bar w}<\delta$ we
have
\[ {\bar w}''+\l {\bar w} \geq (1+O(\delta))w_2 .\]
Let $A$ be the set on which ${\bar w}\geq {\bar w}''$, $B$ the set
on which ${\bar w}<{\bar w}''$. Since for any $x_1 >0$ we have
$\int_x^\infty {\bar w}\geq \int_x^{x+x_1}{\bar w}\geq x_1 {\bar
w}(x+x_1)$, on $A$ we get
\[ \frac{1+\l}{1+O(\d)} {\bar w} \geq \int_x^\infty x_1
{\bar w}(s+x_1)ds\geq x_1 \int_x^{x+x_1} {\bar w}(s+x_1)ds \geq
x_1^2 {\bar w}(x+2x_1).\] In a similar manner, on $B$ we get
\[ \frac{1+\l}{1+O(\d)} {\bar w} \geq x_1^4
{\bar w}(x+4x_1).\]
Thus for all large ${\bar x}$ there is a $k<1$
such that ${\bar w}(x+{\bar x})\leq k {\bar w}(x)$. Let $h(x)={\bar
w}(x)e^{\g x}$, where $\g =\frac{1}{{\bar x}} \ln \frac{1}{k}$. Then
\[h(x+{\bar x})={\bar w}(x+{\bar x})e^{\g x}e^{\g {\bar x}}\leq
{\bar w}(x) e^{\g x}=h(x),\] and thus $h$ is bounded and
$\pi-w(x)=O(e^{-\g x})$ as $x\ra \infty$. Thus for $-\g < {\rm Re}~
\xi <0$ we can define the two-sided Laplace transform of ${\bar w}$
by $W(\xi)=\int_R e^{-\xi x}{\bar w}(x)dx$, and it satisfies
\begin{equation}(\xi^4 +\l \xi^2 -1)W(\xi) =\int_R e^{-\xi x}r({\bar w})dx ,
\label{lt}
\end{equation}
where $r({\bar w})= -\frac{1}{6}{\bar w}^3 +\ldots $. Since the
right side in (\ref{lt}) is defined for $-3\g <{\rm Re}~\xi<0$, by
bootstrap $W$ is analytic in the strip $-m_1 <{\rm Re}~\xi <0$.
Since ${\bar w}$ is a positive decreasing function and
$\int_0^\infty e^{-\xi x}{\bar w}(x)dx =\frac{H(\xi )}{\xi +m_1}$,
where $H$ is analytic in the strip $-m_1 \leq {\rm Re}~\xi <0$, from
Ikehara's Theorem we conclude that $\pi-w(x)=O(e^{-m_1 x})$
\cite{cc}.

Using the method of variation of parameters, ${\bar w}$ satisfies
the integral equation
\begin{eqnarray} \nonumber
{\bar w} (x)&=&c_1 e^{m_1 x}+c_2 e^{m_2 x}+c_3 e^{m_3 x} +c_4 e^{m_4
x}\\\nonumber && +~~a_1 e^{m_1 x} \int_x^\infty
e^{(m_2+m_3+m_4)s}r({\bar w}(s))ds  \\\nonumber &&+~~a_2 e^{m_2 x}
\int_x^\infty e^{(m_1+m_3+m_4)s}r({\bar w}(s))ds\\\nonumber &&
+~~a_3 e^{m_3 x} \int_x^\infty e^{(m_1+m_2+m_4)s}r({\bar w}(s))ds
\\\nonumber && +~~a_4 e^{m_4 x} \int_x^\infty
e^{(m_1+m_2+m_3)s}r({\bar w}(s))ds,
\end{eqnarray}
where $a_i$, $i=1,\ldots,4$, can be calculated expicitly from the
method and necessarily $c_1=c_3=c_4=0$.

Let $c=c_1$, ${\bar w}_0 (z)=c e^{-m_1 z}$, and the sequence ${\bar
w}_n (z)$ be defined by
\begin{eqnarray} \nonumber
{\bar w}_{n+1}(z)=c e^{-m_1 z}  +a_1 e^{m_1 z} \int_z^\infty e^{-m_1
s} r({\bar w}_n (s))ds +a_2 e^{-m_1 z} \int_z^\infty e^{m_1
s}r({\bar w}_n (s))ds \\\nonumber +a_3 e^{m_3 z} \int_z^\infty
e^{-m_3 s}r({\bar w}_n (s))ds +a_4 e^{-m_3 z} \int_z^\infty e^{m_3
s}r({\bar w}_n (s))ds,
\end{eqnarray}
where the integration paths are on the horizontal line $z=x+iy$,
with $y$ fixed. There exists $M$ such that for ${\rm Re}~z \geq M$
and $0\leq {\rm Im}~z \leq \frac{\pi}{m_1}$ we have $|{\bar w}_n
(z)|\leq 2c e^{-m_1 x}$ for all $n$ and ${\bar w}_n (z)$ converges
uniformly to a unique solution ${\bar w}(z)$. Since each ${\bar
w}_n$ is analytic and the convergence is uniform, ${\bar w}$ is also
analytic. Using Picard iterations in which the integrations are on
bounded horizontal segments, ${\bar w}$ is then analytically extended
to the maximal strip of existence $\{ (x,y): x> x_s ,0\leq y \leq
\frac{\pi}{m_1 } \}$.

To show that ${\bar w}(x)$ is odd, first note that ${\tilde
w}(x)=-{\bar w}(-x)$ is also a solution of (\ref{w}). Let $c_0$
correspond to ${\tilde w}(0)=0$. Since the solution of the Picard
iteration with $c=c_0$ is unique, ${\tilde w}={\bar w}$ and ${\bar
w}$ is odd.  $\Box$

To show (\ref{pq}), we first establish various monotonicity
properties of $p$ and $q$. Let $0\leq y \leq \frac{1}{2}
\frac{\pi}{m_1}$. Since ${\bar w}(z)=e^{-m_1 z}[c+o(1)]$, $p<\pi$
and $q>0$ for $x$ large enough. Also, from
\[ \begin{array}{l}
 p_x =m_1 e^{-m_1 x} \cos m_1 y ~~(1+o(1)) ,\\
 q_x =-m_1 e^{-m_1 x} \sin m_1 y ~~(1+o(1)),\\
 p_{xx} =-m_1^2 e^{-m_1 x} \cos m_1 y ~~(1+o(1)) ,\\
 q_{xx} =m_1^2 e^{-m_1 x} \sin m_1 y ~~(1+o(1)),
 \end{array} \]
we conclude that
\[
p_x >0,~~ q_x <0,~~ p_{xx} <0,~~ q_{xx}>0~~{\rm for}~~ x~~ {\rm
large}~~{\rm enough}.
\]
Note that $w=p+iq$ satisfies the system
\begin{equation}  \left\{
\begin{array}{l}
p^{(iv)}+\l p''+\sin p ~~\cosh q=0, \\
q^{(iv)}+\l q''+\cos p ~~\sinh q=0, \end{array} \right. \label{pqs}
\end{equation}
so that
\begin{equation}
\begin{array}{l} p_x =\frac{1}{\sqrt{\l}}\int_x^\infty (1-\cos
\sqrt{\l} (x-s))~~\sin p(s,y)~~ \cosh q(s,y)~~ds ,\\
p_{xx}=\frac{1}{\sqrt{\l}}\int_x^\infty (1-\cos \sqrt{\l}
(x-s)) ~~(\cos p(s,y)~~p_s (s,y) ~~\cosh q(s,y),\\
\hspace{1cm}+ \sin p(s,y) ~~\sinh q(s,y)~~q_s(s,y)) ~~ds,\\
q_x =\frac{1}{\sqrt{\l}}\int_x^\infty (1-\cos \sqrt{\l} (x-s)) ~~\cos
p(s,y)~~\sinh q(s,y)~~ds,\\
\end{array}  \label{pq'}
\end{equation}
Since $|w(z)|\leq 2c e^{-m_1 x}$ uniformly in $y$, from (\ref{pq'})
we see that $p_x >0$, $q_x <0$ and $p_{xx}<0$ hold on $S=\{
(x,y):x\geq \max \{ M, p(\cdot,0)^{-1} (\frac{\pi}{2}) \},0\leq y
\leq \frac{1}{2}\frac{\pi}{m_1}\}$.

Let $\pi -{\bar w}_n =p_n +iq_n$. Since all $p_n$ and $q_n$ satisfy
(\ref{pq}), with $p_n$ in place of $p$ and $q_n$ in place of $q$,
(\ref{pq}) holds on $S$ also for their limits $p$ and $q$.

From (\ref{pq'}) we see that $x_s$ is the value at which $q(x,y)$
becomes infinite as $x \ra x_s +$. From
$p(x,\frac{1}{2}\frac{\pi}{m_1})=\pi$ and the last equation in
(\ref{pq'}), we see that as long as $p_y ,q_y \geq 0$ in
$\{ (x,y): x>x_s , 0\leq y \leq \frac{1}{2}\frac{\pi}{m_1})$,
this singularity will be on the line $y=\frac{1}{2}\frac{\pi}{m_1}$.

Let $S_s=\{ (x,y):x > \max \{ 0, x_s \} ,0< y <
\frac{1}{2}\frac{\pi}{m_1}\}$. Since $q_y =p_x$, from the first
equation in (\ref{pq'}) we see that the first value $x_f$ at which
$p_y$ or $q_y$ is $0$ in $S_s$ can be only such that $p_y =0$.
However, from (\ref{pqs}) we get
\begin{equation} -p_{yyy}+p_y = \int_0^y \sin p ~~\cosh q , \label{mp}
\end{equation}
thus since $0$ would be a minimal value for $p_y$ in $I_f =\{ (x,y):
x=x_f, 0< y < \frac{1}{2}\frac{\pi}{m_1} \}$ and the right side of
(\ref{mp}) is positive, we get a contradiction. Also, since $p_y$ is
harmonic and positive in $S_s$, from the maximum principle we cannot
have $p_y =p_{yy}=0$ on $\{ (x,0):x> \max \{ 0,x_s \} \}$. Thus $p_y ,
q_y >0$ in $S_s$ and $p_{xx} <0$ on $\{ (x,0):x> \max \{ 0,x_s \} \}$.
In a similar way we obtain that $q_{xx}>0$ on $\{ (x,\frac{1}{2}\frac{\pi}{m_1}):
x> \max \{ 0,x_s \} \}$ and $p_{yy}>0$ in $S_s$.

Let $x_s \geq 0$ and $I_s =\{ (x,y): x=x_s, 0< y < \frac{1}{2}\frac{\pi}{m_1}
\}$. In a similar setting, in \cite{jtm} the authors studied the analytic
continuation of $w$ on $I_s$, using Harnack's inequality and the
polynomial form of the nonlinearity. They reached a contradiction and avoided
the eventual analytic continuation to the left-half plane. Such an argument
seems unavailable for a transcendental function, so we argue as in
\cite{am}, adding a few more details for the convenience of the
reader.

Since $p(\cdot,y)$ is decreasing and bounded in $S_s$, let $p(x_s
,y)=\lim_{x\ra x_s +} p(x,y)$. From the last equation in
(\ref{pq'}), we see that there is no $y_s \in
(0,\frac{1}{2}\frac{\pi}{m_1})$ for which $q_x (x,y_s )\ra -\infty$
as $x\ra x_s +$, as then we would have $p_y (x,y)=-q_x (x,y)\ra
\infty$ as $x\ra x_s +$ for all $y\in (y_s
,\frac{1}{2}\frac{\pi}{m_1})$, contradicting that $p$ is bounded.
Let $x_n \ra x_s +$. For any $\d >0$, $p_y (x_n ,y)$ is a sequence
of increasing and bounded functions on
$[0,\frac{1}{2}\frac{\pi}{m_1}-\d]$, so from Helly's Theorem there
is a subsequence converging to an increasing function
$h(y)=\lim_{x_n \ra x_s +}p_y (x_n ,y)$. Passing to the limit on
both sides of $p(x_n ,y)=p(x_n ,0)+\int_0^y p_s (x_n, s)ds$, we get
$p_y (x_s ,y)=h(y)$. Extending in this way to $y\in
[0,\frac{1}{2}\frac{\pi}{m_1})$, we find that $p_y (x_s ,y)$ is
increasing.

Integrating the first equation in (\ref{pqs}) twice, we get
\[ -p_{xx} (x,0) +\l (\pi
-p(x,0))=\int_0^{\frac{1}{2}\frac{\pi}{m_1}} \int_0^y \sin p ~~\cosh
q .\]
Integrating the second equation in (\ref{pqs}) twice, we get
\[ q_{xx} (x,{\scriptstyle\frac{1}{2}\frac{\pi}{m_1}} ) +\l
q(x,{\scriptstyle\frac{1}{2}\frac{\pi}{m_1}} ) -
\frac{\l}{2}\frac{\pi}{m_1} (p_{xxx}(x,0)+p_x (x,0))
=\int_0^{\frac{1}{2}\frac{\pi}{m_1}} \int_0^y \cos p ~~\sinh q .\]
If $p(x_s ,0)\not\equiv 0$, then there are some positive constants
$K_1, K_2$, such that
\[ \lim_{x \ra x_s +} \int_0^{\frac{1}{2}\frac{\pi}{m_1}}
\int_0^y \cos p ~~\sinh q \leq K_1 + \lim_{x \ra x_s +} K_2
\int_0^{\frac{1}{2}\frac{\pi}{m_1}} \int_0^y \sin p ~~\cosh q .\]
Since $q_{xx}(x ,\frac{1}{2}\frac{\pi}{m_1})>0$ and
$q(x ,\frac{1}{2}\frac{\pi}{m_1})\ra \infty$ as $x\ra x_s +$, we get
a contradiction.

To reach (\ref{pq}), it is now enough to rule out the case $x_s <0$.
Since the solution of the initial value problem
\[ \left\{ \begin{array} {l} -p_{yyyy}+ \l p_{yy}=\sin p~~\cosh q,\\
p(0,0)=p_y(0,0)=p_{yy}(0,0)=p_{yyy}(0,0)=0,
\end{array} \right.\]
is $p(0,y)\equiv 0$, it contradicts
$p(0,\frac{1}{2}\frac{\pi}{m_1})=\pi$.  $\Box$


\begin{thebibliography}{99}

\bibitem{as}
     \newblock Yu. M. Aliev and V. P. Silin,
     \newblock \emph{Travelling $4\pi$-kink in nonlocal Josephson
     electrodynamics}, Phys. Lett. A \textbf{177} (1993), 259--262.

\bibitem{am}
     \newblock C. J. Amick and J. B. McLeod,
     \newblock \emph{A singular perturbation problem in needle crystals},
     \newblock Arch. Rational Mech. Anal. \textbf{109} (1990), 139--171.

\bibitem{am4}
     \newblock C. J. Amick and J. B. McLeod,
     \newblock \emph{A singular perturbation problem in water
     waves},
     \newblock Stability Appl. Anal. Contin. Media \textbf{1}
     (1991), 127--148.

\bibitem{at}
     \newblock C. J. Amick and J. F. Toland,
     \newblock \emph{Solitary waves with surface tension. I.
     Trajectories homoclinic to periodic orbits in four dimensions},
     \newblock Arch. Rational Mech. Anal. \textbf{118} (1992),
     37--69.

\bibitem{aekm}
     \newblock G. L. Alfimov, V. M. Eleonsky, N. E. Kulagin and N. V. Mitskevich,
     \newblock \emph{Dynamics of topological solitons in models with nonlocal interactions},
     \newblock Chaos \textbf{3} (1993), 405--414.

\bibitem{bz}
     \newblock P. W. Bates and C. Zhang,
     \newblock \emph{Traveling pulses for the Klein-Gordon equation
     on a lattice or continuum with long-range interaction},
     \newblock Discrete Contin. Dyn. Syst. Ser. A \textbf{16} (2006), 235--252.

\bibitem{cc}
     \newblock J. Carr and A. Chmaj,
     \newblock \emph{Uniqueness of travelling waves for nonlocal
     monostable equations}, Proc. Amer. Math. Soc. \textbf{132},
     2433--2439.

\bibitem{hm}
     \newblock J. M. Hammersley and G. Mazzarino,
     \newblock \emph{A differential equation connected with the
     dendritic growth of crystals},
     \newblock IMA J. Appl. Math. \textbf{42} (1989), 43--75.

\bibitem{in}
     \newblock N. Ishimura and M. Nakamura,
     \newblock \emph{Nonexistence of monotonic solutions of some third-order
     ode relevant to the Kuramoto-Sivashinsky equation},
     \newblock Taiwanese J. Math. \textbf{4} (2000), 621--625.

\bibitem{jtm}
     \newblock J. Jones, W. C. Troy and A. D. MacGillivray,
     \newblock \emph{Steady solutions of the Kuramoto-Sivashinsky
     equation for small wave speed},
     \newblock J. Differential Equations \textbf{96} (1992), 28--55.

\bibitem{pt}
     \newblock L. A. Peletier and W. C. Troy,
     \newblock \emph{Spatial patterns described by the extended
     Fisher-Kolmogorov (EFK) equation: kinks},
     \newblock Differential Integral Equations \textbf{8} (1995),
     1279-1304.

\bibitem{ps}
     \newblock J. K. Perring and T. H. R. Skyrme
     \newblock \emph{A model unified field equation},
     \newblock Nuclear Phys. \textbf{31} (1962), 550--555.

\bibitem{pk}
     \newblock M. Peyrard and M. D. Kruskal,
     \newblock \emph{Kink dynamics in the highly discrete sine-Gordon system},
     \newblock Phys. D \textbf{14} (1984), 88--102.

\bibitem{prg}
     \newblock Y. Pomeau, A. Ramani and B. Grammaticos,
     \newblock \emph{Structural stability of the Korteweg-de Vries
     solitons under a singular perturbation},
     \newblock Phys. D \textbf{31} (1988), 127-134.

\bibitem{s}
     \newblock V. H. Schmidt,
     \newblock \emph{Exact solution in the discrete case for
     solitons propagating in a chain of harmonically coupled
     particles lying in double-minimum potential wells},
     \newblock Phys. Rev. B, \textbf{20} (1979), 4397--4405.

\bibitem{sk}
     \newblock H. Segur and M. D. Kruskal,
     \newblock \emph{Nonexistence of small-amplitude breather
     solutions in $\phi^4$ theory},
     \newblock Phys. Rev. Lett. \textbf{58} (1987), 747--750.

\bibitem{tu}
     \newblock C.-L. Terng and K. Uhlenbeck,
     \newblock \emph{Geometry of solitons},
     \newblock Notices Amer. Math. Soc., \textbf{47} (2000), 17--25.

\bibitem{t}
     \newblock J. F. Toland,
     \newblock \emph{Existence and uniqueness of heteroclinic orbits for the equation
     $\lambda u'''+u'=f(u)$},
     \newblock Proc. Roy. Soc. Edin. A \textbf{109} (1988), 23--36.


\end{thebibliography}
\end{document}